\documentclass[english,letterpaper]{article}
\usepackage[T1]{fontenc}
\usepackage[latin1]{inputenc}
\usepackage{float}
\usepackage{amsmath}
\usepackage{graphicx}
\usepackage{amssymb}
\usepackage{fancyhdr}
\makeatletter
\usepackage{bm}
\usepackage{graphicx,color,psfrag}

\usepackage{babel}
\makeatother

\begin{document}

\title{Dynamical Synapses Causing Self-Organized Criticality in Neural Networks}

\author{A.~Levina$^{1,2,3}$, J.~M.~Herrmann$^{1,4}$, T.~Geisel$^{1,3,4}$\\
{\small $^1$Bernstein Center for Computational Neuroscience G\"{o}ttingen}\\
{\small $^2$Graduate School  ``Identification in Mathematical Models''  }\\
{\small $^3$Max Planck Institute for Dynamics and Self-Organization}\\
{\small $^4$Department of Physics, Georg-August University G\"{o}ttingen}\\
{\small Bunsenstr.~10, 37073 G\"{o}ttingen, Germany}\\[5mm]
}

\maketitle
 \thispagestyle{fancy}

\textbf{Self-organized criticality \cite{Bak_1987} is one of the
key concepts to describe the emergence of complexity in  natural
systems. The concept asserts that a system self-organizes into a critical
state where system observables are distributed according to a power-law.
Prominent examples of self-organized critical dynamics include, e.g.,
 piling of granular media \cite{Frette_1996}, plate 
tectonics~\cite{Gutenberg_1956} and stick--slip motion~\cite{Feder_1991}. 
Critical behavior
 has been shown to bring about optimal computational capabilities~\cite{Maass},
optimal transmission~\cite{Beggs_2003} and storage of 
information~\cite{Haldeman_2005},
and sensitivity to sensory stimuli~\cite{Der,Copelli,Chialvo}. 
In neuronal systems the existence of critical avalanches was predicted in a paper of one 
of the present authors \cite{Eurich_2002} and observed experimentally by 
Beggs and Plenz \cite{Beggs_2003,Beggs_2004,Plenz2007}. Nevertheless, while in the 
experiments critical avalanches were found generically in the sense of genuine 
self-organized criticality, in the model of Ref.~\cite{Eurich_2002} they only show 
up, if the set of parameters is fine-tuned externally to a critical transition 
state. In the present paper we demonstrate analytically 
and numerically that by assuming (biologically more
realistic) dynamical synapses~\cite{Markram_1996} in a spiking neural network, the neuronal
avalanches turn from an exceptional phenomenon into a typical and robust
self-organized critical behavior 
if the total resources of neurotransmitter are sufficiently large.
}

\begin{figure}
\begin{center}
\includegraphics[width=0.56\columnwidth]{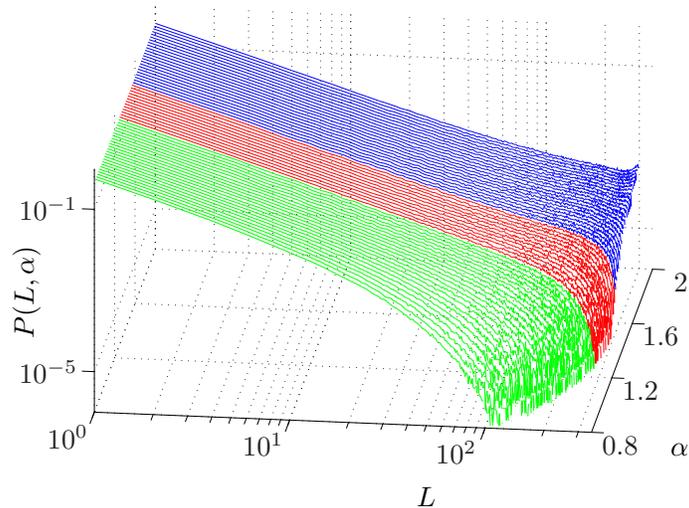}
\end{center}
\caption{Distribution of avalanche sizes for different coupling strengths $\alpha$.
At $\alpha<1.3$ small avalanches are preferred yielding a subcritical distribution. 
The range of connectivity parameters near $\alpha=1.4$ appears critical. 
For $\alpha>1.6$ the distribution
is supercritical, i.e.~a substantial fraction of firing events spreads
through the whole system. These results are shown for 
$N=300,\:\nu=10,\; u=0.2,\; I^{{\rm {ext}}}=0.025$,
other parameter values are discussed below.\label{Fig:distributions}}
\end{figure}

In multi-electrode recordings  on  slices
of rat cortex and neuronal cultures~\cite{Beggs_2003,Beggs_2004}
neuronal avalanche were observed whose sizes were distributed according to a power-law with
an exponent of -3/2. The distribution was stable over a long period
of time.  Variations of the dynamical behavior are induced by application or
wash-out of neuromodulators. Qualitatively identical behavior can be reached
in models like \cite{Eurich_2002,Fukai2006} by variations of a global
connectivity parameter. In these models, criticality only shows up, if the interactions are fixed precisely at a specific value or connectivity structure.

Here we study a model with activity-dependent depressive synapses and show that  existence of several dynamical regimes can be reconciled
with parameter-independent criticality.
We find that  synaptic depression causes the mean synaptic strengths to approach a critical
value for a certain range of interaction parameters, while outside
this range other dynamical behaviors are prevalent, cf.~Fig.~\ref{Fig:distributions}.
We analytically derive  an expression for the average coupling strengths
among neurons and the average inter-spike intervals in a mean-field approach.
The mean field approximation is applicable here even in the critical
state, because the quantities that are averaged  do not exhibit power-laws,
but unimodal distributions. These mean values obey a 
self-consistency equation which allows us to identify the mechanism that
drives the dynamics of the system towards the critical regime. Moreover, 
the critical regime induced by the synaptic dynamics is robust to parameter changes. 

Consider a network of $N$ integrate-and-fire neurons. Each neuron
is characterized by a membrane potential $0<h_{i}(t)<\theta$. The
neurons receive external inputs by a random process $\xi_{\tau}\in\left\{ 1,\dots,N\right\} $
which selects a neuron $\xi_{\tau}\left(t\right)=i$ at a rate $\tau$ and
advances the membrane potential $h_{i}$ by an amount $I^{\textrm{ext}}$.
Each neuron integrates inputs until it reaches a threshold $\theta$.
As soon as $h_{i}(t)>\theta$ the neuron emits a spike which is delivered
to all postsynaptic neurons at a fixed delay $\tau_{d}$. 
The membrane potential is reset by 
$h_{i}(t_{\textrm{sp}}^{+})=h_{i}(t_{\textrm{sp}})-\theta$. 
For simplicity we will assume in the following that $\theta=1$.
Super-threshold activity is communicated along neural
connections of a strength proportional to $J_{ij}$ to other neurons 
and may cause them to fire. In this way an avalanche of neural
activity of size $L\ge1$ is triggered. 
More precisely, an avalanche is a period of activity that is initiated by 
an external input and that terminates when no further neuron becomes activated. 
We define the size of the avalanche as the number of participating neurons.
The dynamics of the membrane potential is described by the following equation
\begin{equation}
\dot{h}_{i}=\delta_{i,\xi_{\tau}(t)}I^{\textrm{{ext}}}+\frac{1}{N}\sum_{j=1}^{N}uJ_{ij}\delta\left(t-t_{\textrm{{sp}}}^{j}-\tau_{d}\right)\label{eq:membrane_pot}.\end{equation}

In studies of self-organized criticality typically a separation of time scales
is assumed which enters in Eq.~\ref{eq:membrane_pot} by the condition 
$\tau_{d}\ll\tau$. It allows us to assume that external input is absent during 
avalanches.  Later, in the discrete version of
the model, $\tau$ will play the role of the temporal step size.
The variables $J_{ij}$ are subject to the dynamics 
\begin{equation}
\dot{J}_{ij}=\frac{1}{\tau_{J}}\left(\frac{\alpha}{u}-J_{ij}\right)-
uJ_{ij}\delta\left(t-t_{\textrm{{sp}}}^{j}\right),
\label{eq:j_dyn}\end{equation}
which describes the amount of available neurotransmitter in the corresponding
synapse~\cite{Markram_1996}. 
Namely, if a spike arrives at the synapse, the available transmitter
is diminished by a fraction $u$, i.e.~the synaptic strength decreases
due to the usage of transmitter resources.  
If the presynaptic neuron is silent then the
synapse recovers and and its strength $J_{ij}$ 
approaches the value $\alpha/ u$ at a slow time scale $\tau_{J}=\tau \nu N$ 
with $1<\nu\ll N$. 
Thus, the maximal strength of a connection is determined by the parameter $\alpha$
and can be observed only when the synapse is fully recovered. The behavior of the
network is determined, however, by the averaged synaptic strength which will be
denoted by $\left\langle J_{ij}\right\rangle$ with the average taken
with respect to the distribution of inter-spike intervals. In order
to obtain our main result we will calculate this effective value and 
use it in a static network. The uniform strengths of the static network
are denoted by $\alpha_0$.

\begin{figure}
\begin{center}\includegraphics[width=0.5\columnwidth]{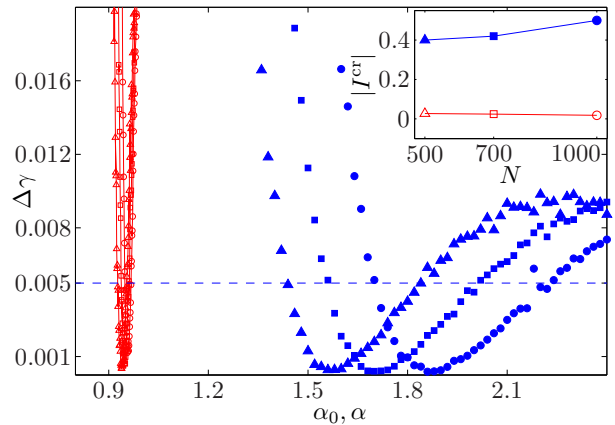}\end{center}
\caption{The range of connectivity parameters which cause a critical dynamics 
extends with system size. The mean-squared deviation from the best-matching 
power-law is plotted  
in dependence of the connection strengths $\alpha_0$ and  $\alpha$ for
static synapses and depressive synapses, respectively. Blue circles, squares and
triangles stand for networks with dynamical synapses and system sizes $N=1000,\:700$,
and $500$, respectively. Red symbols represent
the static model. Note that the minimum of all curves depends of the network size 
\cite{Eurich_2002}.  The inset (same symbols) shows the lengths of
the parameter intervals where the deviation from the best-matching power-law 
is smaller than an ad-hoc threshold ($\Delta \gamma=0.005$). 
\label{Fig:Critical_Region}}
\end{figure}

If the external drive and the synaptic weights are small, the activity
of the network consists of short burst-like events which
are initiated by a particular external input. The firing events are
separated by relatively long relaxation intervals when external
inputs are integrated. We may thus be tempted to assume $J\approx\frac{\alpha}{u}$
before any spiking event. In general, however, we must take into account
that the efficacy of a synapse varies in a usage-dependent way which compensates
large levels of network activity. 
Depending on the synaptic strength the network can produce a rich
repertoire of behaviors. In Fig.~\ref{Fig:distributions}, we show
examples of avalanche size distributions for various values of $\alpha$.
For small values of $\alpha$, subcritical avalanche-size distributions
are observed. This regime is characterized by a negligible number
of avalanches that extend to the system size. Near $\alpha_{\textrm{cr}}$
the system has an approximate power-law avalanche distribution for
avalanche sizes almost up to the system size where an exponential
cut-off is observed. The mean-squared deviation from an exact power law
is shown in Fig.~\ref{Fig:Critical_Region}.
Finally, the distribution of avalanche sizes becomes non-monotonous
when $\alpha$ is well above the critical value $\alpha_{\textrm{{cr}}}$.

\begin{figure}
\begin{center}
\includegraphics[width=0.5\columnwidth]{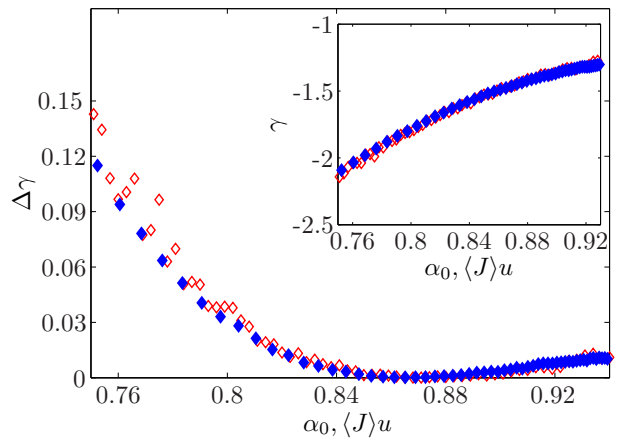} 
\end{center}
\caption{Rescaling of depressive synapses. 
A network with static synapses of uniform strength $\alpha_0$ has the same 
statistical properties as a network with dynamic synapses if $\alpha_0$
is fixed at the average synaptic strength of the dynamical case, 
i.e.~if $\alpha_0:=\langle J\rangle u$. 
The mean-squared deviation $\Delta \gamma$  
from the best-matching power-law is shown  
as a function of the synaptic strength $\alpha_0$ for static synapses (red symbols) 
and the mean synaptic strength for dynamical synapses (blue symbols), respectively.
The inset (same symbols) shows the exponent $\gamma$ of the best-matching power-law 
in the two cases. 
Parameters are $N=100$, $\nu=10$, $u=0.2$. \label{Fig:Slope_mean_syn}}
\end{figure}

In preliminary numerical studies we had assumed a model with facilitating and depressing synapses  \cite{Levina}. Here we conclude that facilitating synapses are not necessary 
to evoke self-organized critical avalanches in spiking neural networks, 
depressing synapses are sufficient. This is in line with
the observation \cite{Markram_1998} that synapses that connect excitatory neurons are
largely depressive. 
To identify the parameters of the avalanche size distribution it is sufficient 
to determine the average synaptic strength: As seen in Fig.~\ref{Fig:Slope_mean_syn}
both the power-law exponent and
the mean-squared deviation from the power-law are the same for networks
with dynamical synapses and networks with static synapses if the strength of the static
synapses is chosen as $\alpha_0=u\left\langle J_{ij}\right\rangle$. 
In order to calculate the average synaptic strength analytically,
we consider in addition the neural inter-spike intervals $\Delta^{\textrm{isi}}$. 
On the one hand, if the inter-spike intervals are short then the synapses have 
a short time to recover and the average synaptic strength resides at a low level. 
On the other hand, large synaptic strengths lead to long avalanches 
and to large input to neurons during the avalanches, which tends to shorten
the inter-spike intervals. 

\begin{figure}
\begin{center}
\includegraphics[width=0.6\columnwidth]{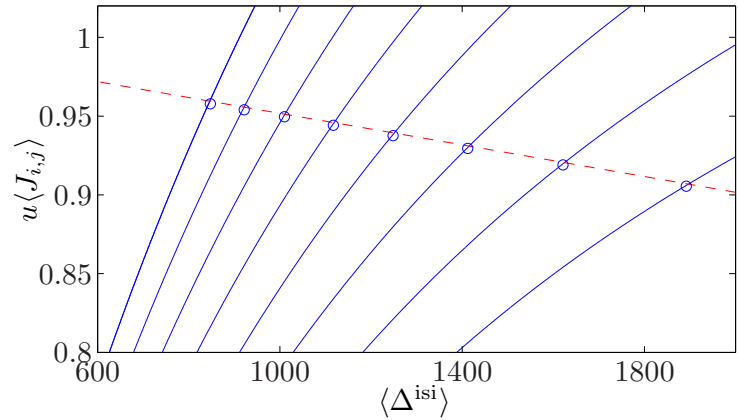}
\end{center}
\caption{The graphical solution of Eqs.~\ref{eq:self_Jij} establishes
a functional relation between the average synaptic strength and inter-spike 
interval. It is obtained by
the intersections of the solutions of Eq.~\ref{eq: ISI(a)}
(dashed line) and Eq.~\ref{eq: J(IAI)} (solid lines) for
$\alpha=1.3,\ldots,2.0\;$ in steps of $0.1$ (from right to left).
This solution agrees well with the results of simulations  (circles) 
of a network with the same values of $\alpha$. 
Parameters are $N=500,\:$ $\nu=10,\: u=0.2$.  \label{Fig:ISI_J}}
\end{figure}

This trade-off determines the expected values of the synaptic strengths and the 
inter-spike intervals which are realized by the dynamics of the network. 
In order to express this reasoning more formally, we
solve the dynamical equations (\ref{eq:membrane_pot}) and (\ref{eq:j_dyn})
 based on a stationarity assumption for both the synaptic strengths and the 
 inter-spike interval. Neither of these quantities has a power-law distribution
and their first moments exist. In \textbf{Methods} we derive expressions 
of the mean synaptic strength $\left\langle J_{ij}\right\rangle$ and the
mean value of the inter-spike intervals distribution $\langle\Delta^{\mathrm{isi}}\rangle$.
The stochastic dependency of the two quantities is reflected in 
a mutual dependence of their averages.  Each of the dependencies is derived analytically
which allows us to formulate the self-consistency of the stationarity requirement as
the simultaneous solution of the mean-field equations
\begin{equation}
\left\langle J_{ij}\right\rangle=
\left\langle J_{ij}\right\rangle
\left(\rule{0cm}{0.4cm} \left\langle \Delta^{\mathrm{isi}}\right\rangle \right)
\hskip 8mm \hbox{and} \hskip 8mm
\left\langle \Delta^{\mathrm{isi}} \right\rangle=
\left\langle \Delta^{\mathrm{isi}} \right\rangle
\left(\rule{0cm}{0.4cm} \left\langle J_{ij} \right\rangle \right)
\label{eq:self_Jij}
\end{equation}
which can be determined graphically from the intersections of the solutions 
of Eqs.~\ref{eq: J(IAI)} and \ref{eq: ISI(a)}, cf.~Fig.~\ref{Fig:ISI_J}.
The mean-field solution is confirmed by direct network simulations that are
are represented by the circles in~Fig.~\ref{Fig:ISI_J}.
The solution is unique for any $\alpha$. The stationary
distribution is less sensitive to  changes of the parameter $\alpha$
near the critical value of the synaptic strength than further away from it.
This brings about the large critical region for the model with depressive synapses, 
cf.~Fig.~\ref{Fig:Critical_Region}. 

Furthermore we want to discuss the stability of the solution of the self-consistency 
equation (\ref{eq:self_Jij}). If we apply a perturbation
$\Delta J$ to all synapses at  time $t_{p}$ such that for 
each $i, j$ $\widetilde{J_{ij}}=J_{ij}+\Delta J$,
we can show with some simple computations,  that before the next spike 
the synaptic strengths are on average smaller than
$\widetilde{J_{ij}}$. In the simulated system
the average synaptic strength is driven back to the fixed point by a few spikes, 
such that the solution of (\ref{eq:self_Jij}) is indeed stable for the critical state. 

Both the numerical study in Ref.~\cite{Levina} and the analysis
presented so far refer to finite systems. 
In order to check whether the trend that is visible 
in Fig.~\ref{Fig:Critical_Region} continues for
larger network sizes, we consider the behavior of the mean synaptic 
strength in the thermodynamic
limit $N\to\infty$ we compute the expectation value of the 
avalanche-size distribution (\ref{eq:EEH_formula}),
$\langle L\rangle=\frac{N}{N-(N-1)\alpha_{0}}$~\cite{Eurich_2002},
and insert it into the self-consistency equation (\ref{eq:self_Jij})
\begin{equation}
\frac{\alpha\left(1-e^{-\frac{1}{\nu N}\langle\Delta^{\mathrm{isi}}\rangle}\right)}
{1-(1-u)e^{-\frac{1}{\nu N}\langle\Delta^{\mathrm{isi}}\rangle}}
=\frac{N}{N-1}-\frac{I^{{\rm ext}}\langle\Delta^{\mathrm{isi}}\rangle}{N-1}.
\label{eq:Self_Const_Limit}
\end{equation}
  In the limit $N\to\infty$ we should scale the external input $I^{{\rm ext}}\sim N^{-w}$ and $w>0$. We now distinguish the following cases. 
a) If $\langle\Delta^{\mathrm{isi}}\rangle\sim N^{1+\epsilon}$ and $\epsilon>w$
then the right hand side (r.h.s.) of (\ref{eq:Self_Const_Limit}) tends to $-\infty,$
while the l.h.s. is always larger than $0$. 
b) If $\langle\Delta^{\mathrm{isi}}\rangle\sim N^{1+\epsilon}$ and $0<\epsilon\leq w$
then the r.h.s. of (\ref{eq:Self_Const_Limit}) tends to $1$  (or 0 if $\epsilon= w$) while the
l.h.s. $\alpha,$ hence a solution is only possible if
$\alpha=1$ and in this case $u\langle J_{ij}\rangle\to1$.
 c) If $\langle\Delta^{\mathrm{isi}}\rangle\sim N^{1-\epsilon}$ and $\epsilon<0$
then the r.h.s. of (\ref{eq:Self_Const_Limit}) tends to $1,$ while the
l.h.s. approaches 0.
 d) If $\langle\Delta^{\mathrm{isi}}\rangle\sim N,$
 we can assume that $\langle\Delta^{\mathrm{isi}}\rangle=cN+o(N)$.
From (\ref{eq:Self_Const_Limit}) for $\alpha>1$ we can find the
unique solution $c=-\nu\left(\ln(\alpha-1)-\ln(\alpha-1+u)\right)$.
In all cases when the solution exists, $u\langle J_{ij}\rangle\to1,$ which
we know to be the  critical connectivity for the network with statical synapses
in the limit $N\to\infty$. Hence in the thermodynamical limit the
network with dynamical synapses becomes critical for any $\alpha\geq1$.

In this paper we have focused on fully connected networks and neurons without leak currents for reasons
of analytical tractability.  We now discuss the results of various generalizations which we have investigated numerically.
If the network has only partial connectivity, the results stay the same, if the synaptic
strengths are properly rescaled. In a random network of size
$N$ with connectivity probability $c$, the critical parameter 
$\alpha$ is approximately equal to $\alpha_{N}^{{\rm {cr}}}/c$,
where $\alpha_{N}^{{\rm {cr}}}$ is obtained from the critical parameter region
of the fully connected network of size $c \times N$. 
If the connections in a partially connected random network are not chosen independently 
(e.g.~``small-world'' connectivity~\cite{Chen_2005}) one finds even more accurate power-laws
than for the independent case with the same average connectivity.
A similar phenomenon occurs in the grid network \cite{Arcangelis06} which has been used to model criticality in  EEG recordings.

If in Eq.~\ref{eq:membrane_pot} we add
a leak term, which is present in biologically realistic situations
\begin{equation}
\dot{h}(t)= - \tau_{l}^{-1} h(t) + C + \delta_{i,\xi_{\tau}(t)}I^{\textrm{{ext}}}+\frac{1}{N}\sum_{j=1}^{N}uJ_{ij}\delta\left(t-t_{\textrm{{sp}}}^{j}-\tau_{d}\right),
\label{leak_eq}
\end{equation}
we find numerically that
the distribution of the avalanche sizes remains a power law for leak time-constants
up to $\tau_l \approx 40 ms$. In (\ref{leak_eq}) we included a constant compensatory 
synaptic current $C$ which depends on $\tau_l$ and summarizes neuronal self-regulatory 
mechanisms. In this way the probability of the neuron to stay near threshold
is conserved and avalanches are triggered in a similar way as in the non-leaky case.
The resulting power-law exponent is slightly smaller than $-1.5$ and 
reaches values close to $-2$ for strong leakage in simulations of a network of $N=500$ 
neurons.

In summary, we have presented an analytical and numerical study of spiking
neural networks with dynamical synapses. Activity-dependent 
synaptic regulation leads to the self-organization of the
network towards a critical state.  
The analysis demonstrates that 
mean synaptic efficacy hereby plays a crucial role.  We 
explained how the critical state depends on the maximally available 
resources and have shown
that in the thermodynamic limit the network becomes critical for any
$\alpha\geq1$, i.e.~that criticality is an intrinsic phenomenon produced by the
synaptic dynamics. The mean field quantities are in very good agreement
with simulations and were shown to be robust with respect to perturbations 
of the model parameters.

%
\paragraph*{Acknowledgments: }

We thank Manfred Denker and Tomoki Fukai for useful discussions. 
This work was partially supported by the BMBF in the framework of
the Bernstein Centers for Computational Neuroscience,
grant number {01GQ0432}.
A.L. has received support from DFG Graduiertenkolleg N1023.
%

\section*{Methods}

In this section we give an explicite derivation of the self-consistency relation 
(\ref{eq:self_Jij}). 
Solving Eq.~\ref{eq:j_dyn}  between two spikes of neuron $j$ we find 
\begin{equation}
J_{ij}(t_2^-)=\frac{\alpha}{u}\left(1-\left(1-\frac{u}{\alpha}J_{ij}(t_1^+)\right)
e^{-\left(t_{2}-t_{1}\right)/\tau_{J}}\right), \label{eq: <J1>}
\end{equation}
where the synaptic strengths immediately before and after a spike of neuron $j$ 
at time $t_{s}$ are denoted by $J_{ij}(t_s^-)$ and $J_{ij}(t_s^+)$, respectively.
Within a short interval containing the spike,
$J_{ij}$  decreases by a fraction $u$ such that 
$J_{ij}(t_1^+)=\left(1-u\right)J_{ij}(t_1^-)$.

The average synaptic strength  $\left\langle J_{ij}\right\rangle$
is the expectation value of $J_{ij}(t_s^-)$. Analogously, 
$\left\langle \Delta^{\mathrm{isi}} \right\rangle$ refers to the inter-spike
interval $\Delta^{\mathrm{isi}}$. The random variables $J_{ij}(t_s^-)$ and 
$\Delta^{\mathrm{isi}}$ both depend on the distribution of external inputs
and are thus related. The self-consistency conditions (\ref{eq:self_Jij})
express this relation for the respective averages.
%
Assuming that $\left\langle J_{ij}\right\rangle$ depends essentially only
on the mean inter-spike interval, we obtain 
from (\ref{eq: <J1>}):
\begin{equation}
\langle J_{ij}\rangle=
\frac{\alpha}{u}\frac{1-e^{-\frac{1}{\nu N}\langle\Delta^{\mathrm{i
si}}\rangle}}{1-(1-u)e^{-\frac{1}{\nu
 N}\langle\Delta^{\mathrm{isi}}\rangle}},\label{eq: J(IAI)}\end{equation}
where the average is performed in discrete time with step width $\tau$, 
i.e.~$\tau_{J}=\nu N$.

We are now going to establish a relation between $P\left(\Delta^{\mathrm{isi}}\right)$
and the inter-avalanche interval distribution $Q\left(\Delta^{\mathrm{iai}}\right)$.
It can be shown  that the neuronal membrane
potentials  before an avalanche are uniformly distributed on the
interval $[\epsilon_{N},\theta]$, where $\epsilon_{N}$ is a lower bound of $h_{i}(t_{\textrm{sp}})-\theta$
with $\epsilon_{N}\to0$ for $N\to\infty$.   Under
these conditions, $Q(\Delta^{\mathrm{iai}})$ has a geometric
distribution\begin{equation}
Q\left(\Delta^{\mathrm{iai}}\right)=\frac{I^{\mathrm{\mathrm{ext}}}}{\theta-\epsilon_{N}}\left(1-\frac{I^{\mathrm{\mathrm{ext}}}}{\theta-\epsilon_{N}}\right)^{\Delta^{\mathrm{iai}}}.\label{eq:Q_IAI}\end{equation}
Let $k_j$ be the number of avalanches between two spikes of the neuron $j$. A mean-field approximation relates the averages of the distributions
of inter-spike and inter-avalanche intervals
\begin{equation}
\left\langle \Delta^{\mathrm{isi}}\right\rangle =\langle k\rangle\langle\Delta^{\mathrm{iai}}\rangle.\label{eq:isi_iai_means}\end{equation}
The average inter-avalanche interval is easily computed from (\ref{eq:Q_IAI})
\begin{equation}
\langle\Delta^{\mathrm{iai}}\rangle=\frac{\theta-\epsilon_{N}}{I^{\mathrm{\mathrm{ext}}}}.\label{eq:IAI_mean}
\end{equation}
In order to calculate the average number of avalanches between two
spikes of a neuron, we compute the time to reach the threshold by accumulating
external inputs and spikes from other neurons during avalanches, i.e.
\begin{equation*}
\langle\kappa\rangle=\frac{\theta}{u\left\langle J_{ij}\right\rangle \left\langle L\right\rangle +I_{\textrm{ext}}\frac{\langle\Delta^{\mathrm{iai}}\rangle}{N}},
\end{equation*}
 where $\left\langle L\right\rangle $ is the mean avalanche size
and $\frac{1}{N}$ is the probability that neuron $j$ is receiving an
input. The distribution of avalanche sizes can be computed analytically
for a network with static synapses of strength $\alpha_{0}$ \cite{Eurich_2002}:
\begin{equation}
P(L,\alpha_{0},N)  =  L^{L-2}{\binom{N-1}{L-1}}\left(\frac{\alpha_{0}}{N}\right)^{L-1} 
 \left(1-L\frac{\alpha_{0}}{N}\right)^{N-L-1}  \frac{N(1-\alpha_{0})}{N-(N-1)\alpha_{0}}.\label{eq:EEH_formula}
\end{equation}
In the case of dynamical synapses we apply a mean field approximation and set 
$\alpha_{0}=u\left\langle J_{ij}\right\rangle $
in (\ref{eq:EEH_formula}). This allows us to  
compute  $\langle L\rangle$ as a function of $\left(u\left\langle J_{ij}\right\rangle \right)$. 

Combining (\ref{eq:isi_iai_means}), (\ref{eq:IAI_mean}), and (\ref{eq:EEH_formula})
we obtain a relation between the inter-spike interval and the average synaptic
strength.\begin{equation}
\left\langle \Delta^{\mathrm{isi}}\right\rangle =\frac{\theta-\epsilon_{N}}
{I^{\mathrm{\mathrm{ext}}}}\frac{\theta}{u\left\langle J_{ij}\right\rangle 
\left\langle L\right\rangle
\left(u\left\langle J_{ij}\right\rangle \right)+I_{\textrm{ext}}\frac{\theta-\epsilon_{N}
}{N}}.\label{eq: ISI(a)}\end{equation}
The self-consistency equations (\ref{eq:self_Jij}) arise from (\ref{eq: J(IAI)}) and
(\ref{eq: ISI(a)}). Its solution is obtained by numerical analysis of 
the two independent relations (\ref{eq: J(IAI)}) and (\ref{eq: ISI(a)}), cf.~Fig.~\ref{Fig:ISI_J}.

\end{document}